\def\Roman#1{\uppercase\expandafter{\romannumeral#1}}
\documentclass[a4paper,12pt]{article}
\usepackage{cite}
\usepackage{amssymb,amsfonts}
\usepackage{amsmath}
\usepackage[mathscr]{eucal}
\usepackage{mathrsfs}
\usepackage[utf8x]{inputenc}

\title{Non-zero momentum level reduction in  path integrals for  dynamical systems with symmetry  given on a product manifold 
consisting of
the total space of the principal fiber bundle and a vector space}
\author{S. N. Storchak\\
\small{ A. A. Logunov Institute for High Energy Physics}\\
\small{of NRC ``Kurchatov Institute'',}\\
\small{Protvino, 142281, Russian Federation,}\\
}

\begin{document}

  \maketitle

\begin{abstract}
The case of non-zero momentum level reduction in  Wiener path integrals for a mechanical system with symmetry  describing   the motion  of two scalar particles with interaction on a Riemannian product manifold with the given action a compact semisimple Lie group  is considered. The original product manifold consists of the vector space and a smooth compact finite-dimensional Riemannian manifold, which, due to the action of the group, can be regarded as the total space of the principal fiber bundle.
The integral relation between the path integrals representing the fundamental solutions of the backward Kolmogorov equation defined on the total space of the principal fiber bundle (the original Riemannian  product manifold) and the corresponding backward Kolmogorov equation gion the space of the sections of the associated covector bundle is obtained.

\begin{flushleft}{\bf{KeyWords:}} Marsden-Weinstein reduction, Kaluza-Klein theories, Path integral, Stochastic analysis.\\
{\bf{MSC:}} 81S40 53B21 58J65
\end{flushleft}
\end{abstract}

\section{Introduction}
This work can be considered as a supplement to our previous work \cite{Storchak_2019}, where the reduction procedure in path integrals for a special dynamical system was considered. The interest in studying this system is due to the fact that it can serve as a model for describing the interaction of the gauge fields with the scalar fields. In our previous work, we considered the  case of the  zero momentum level of the Marsden-Weinsten  reduction\cite{AbrMarsd} in the original mechanical system.  Here we will study the general case of the reduction onto the  non-zero momentum level. In the path integrals, the reduction procedure is performed 

The path integral reduction is based on two transformation of the path integrals. The first transformation is the factorization of the path integral measure which can be done by using the non-linear filtering equation from the stochastic process theory. The second transformation of the path integrals is due to the Girsanov transformation of the stochastic processes. In the present work we will deal with the  Girsanov transformation since the first part of the path integral transformation was already studied in \cite{Storchak_2019}.

\section{Definition}
Let us briefly recall the main definitions used in our previous works.

We  are interested in   the backward Kolmogorov equation which is given on a smooth compact Riemannian manifold $ \tilde {\mathcal P} = \mathcal P \times \mathcal V $:
\begin{equation}
\left\{
\begin{array}{l}
\displaystyle
\left(
\frac \partial {\partial t_a}+\frac 12\mu ^2\kappa \bigl[\triangle
_{\cal P}(p_a)+\triangle
_{\cal V}(v_a)\bigr]+\frac
1{\mu ^2\kappa m}V(p_a,v_a)\right){\psi}_{t_b} (p_a,t_a)=0,\\
{\psi}_{t_b} (p_b,v_b,t_b)=\phi _0(p_b,v_b),
\qquad\qquad\qquad\qquad\qquad (t_{b}>t_{a}),
\end{array}\right.
\label{1}
\end{equation}
where $\mu ^2=\frac \hbar m$ , $\kappa $  is a real positive
parameter,  $V(p,f)$ is the group-invariant potential term:
 $V(pg,g^{-1}v)=V(p,v)$, $g\in \mathcal G$, 
$\triangle _{\cal P}(p_a)$ is the Laplace--Beltrami operator on a
manifold $\cal P$ and $\triangle
_{\cal V}(v)$ is the Laplacian on the vector space $\cal V$. Locally, in the  coordinates $(Q^A,f^a)$ of the point $(p,v)$\footnote{In  our formulas  we  assume that there is 
sum over the repeated indices. The indices denoted by the capital
letters  ranged from 1 to $n_{\cal P}=\rm{dim} \cal P$, and the small Latin letters, except $i,j,k,l$, -- from 1 to $n^{\cal V}=\dim \cal V$.} the Laplacian is given by
\begin{equation}
\triangle _{\cal P}(Q)=G^{-1/2}(Q)\frac \partial {\partial
Q^A}G^{AB}(Q)
G^{1/2}(Q)\frac\partial {\partial Q^B},
\label{2}
\end{equation}
where $G=det (G_{AB})$ and $G_{AB}$ are the components of the initial Riemannian
metric given  in the coordinate basis $\{\frac{\partial}{\partial
Q^A}\}$.

 $\triangle_{\cal V}$ is given by 
\[
 \triangle_{\cal V}(f)=G^{ab}\frac{\partial}{\partial f^a\partial f^b},
\]
By the assumption used in the paper, the matrix $G_{ab}$ consists of fixed constant elements. It is admitted that $ G_{ab}$ may have off-diagonal elements.

In case of fulfillment of  smooth
requirements imposed on 
 the coefficients and the initial function of
equation (\ref{1}), the solution of  equation, as it follows from
\cite{Dalecky_1}, can be represented as follows:
\begin{eqnarray}
{\psi}_{t_b} (p_a,v_a,t_a)&=&{\rm E}\Bigl[\phi _0(\eta_1 (t_b),\eta_2(t_b))\exp \{\frac
1{\mu
^2\kappa m}\int_{t_a}^{t_b}V(\eta_1(u),\eta_2(u))du\}\Bigr]\nonumber\\
&=&\int_{\Omega _{-}}d\mu ^\eta (\omega )\phi _0(\eta (t_b))\exp
\{\ldots 
\},
\label{orig_path_int}
\end{eqnarray}
where ${\eta}(t)$ is a global stochastic process on a manifold 
$\tilde{\cal P}=\cal P\times \cal V$, formed from the processes $\eta_1(t)$ and $\eta_2(t)$; ${\mu}^{\eta}$ is the path integral measure  on
the path space $\Omega _{-}=\{\omega (t)=\omega^1(t)\times\omega^2(t):\omega^{1,2} (t_a)=0, \eta_1
(t)=p_a+\omega^1 (t),\eta_2(t)=v_a+\omega^2(t)\}$ given on manifold $\tilde {\mathcal P}$.

In a local chart $(U_{\cal P}\times U_{\cal V},\varphi )$ of the manifold $\tilde {\mathcal P}$, the process $\eta (t)$ is given by the solution of two stochasic differential equations:
\begin{equation}
d\eta_1^A(t)=\frac12\mu ^2\kappa G^{-1/2}\frac \partial {\partial
Q^B}(G^{1/2}G^{AB})dt+\mu \sqrt{\kappa }{\cal X}_{\bar{M}}^A(\eta_1
(t))dw^{
\bar{M}}(t),\\
\label{eta_1}
\end{equation}
and
\begin{equation}
 d\eta_2^a(t)=\mu \sqrt{\kappa }{\cal X}_{\bar{a}}^b 
dw^{
\bar{b}}(t)\\
\label{eta_2}
\end{equation}
(${\cal X}_{\bar{M}}^A$ and ${\cal X}_{\bar{a}}^b$ are  defined  by the local equalities
$\sum^{n_{\mathcal P}}_{\bar{\scriptscriptstyle K}\scriptscriptstyle =1}{\cal
X}_{\bar{K}}^A{\cal X}_{\bar{K}}^B=G^{AB}$ and  $\sum^{n_{\mathcal V}}_{\bar{\scriptscriptstyle a}\scriptscriptstyle =1}{\cal
X}_{\bar{a}}^b{\cal X}_{\bar{a}}^c=G^{bc}$, $dw^{\bar{M}}(t)$ and $dw^{\bar{b}}(t)$ are the independent Wiener processes.
Here and what follows  we  denote the Euclidean
indices by over-barred indices).

 The global semigroup determined by equation (\ref{orig_path_int})  is defined in \cite{Dalecky_1, Dalecky_2} by the limit (under the refinement of the time interval) of the superposition of the local semigroups. In our case it is given by
\begin{equation}
\!\psi _{t_b}(p_a,v_a,t_a)=U(t_b,t_a)\phi _0(p_a,v_a)=
{\lim}_q {\tilde U}_{\eta}(t_a,t_1)\cdot\ldots\cdot
{\tilde U}_{\eta}(t_{n-1},t_b)
\phi _0(p_a,v_a),
\label{6}
\end{equation}
where  each of the local semigroup  
${\tilde U}_{\eta}$ is as follows:
\[
 {\tilde U}_{\eta}(s,t) \phi (p,v)={\rm E}_{s,p,v}\phi (\eta_1
 (t),\eta_2(t))\,\,\,\,\,\,
 s\leq t\,\,\,\,\,\,\eta_1 (s)=p,\;\eta_2 (s)=v.
 \]
These local semigroups are also given by the path integrals with the integration measures determined by the local representatives
 $\varphi
 ^{\tilde{\cal P}}(\eta_t)
\equiv\{\eta ^A_1 (t),\eta^a_2(t)\}$ of the global stochastic process $\eta(t)$.

On the Riemannian manifold $\tilde{\mathcal P}=\mathcal P \times \mathcal V$, we are given a smooth isometric free and proper action of a compact semisimple Lie group $\mathcal G$: $(\tilde p,\tilde v)=(p,v)g=(pg,g^{-1}v)$. In a local coordinates $(Q^A,f^a)$, this action is given as follows:
\[
 {\tilde Q}^A=F^A(Q,g),\;\;\;\;{\tilde f}^b=\bar D^b_a(g)f^a,
\]
where $\bar D^b_a(g)\equiv D^b_a(g^{-1})$,
and by $D^b_a(g)$ we denote the matrix of  the finite-dimensional representation of the group $\mathcal G$
acting on the vector space $\mathcal V$.

 The Riemannian metric of the manifold $\tilde{\mathcal P}$ can be written as follows:
\begin{equation}
 ds^2=G_{AB}(Q)dQ^AdQ^B+G_{ab}\,df^adf^b.
\label{metr_orig}
\end{equation}

The Killing vector fields for the  metric (\ref{metr_orig}) given on the  manifold $\tilde{\mathcal P}$ are   the  vector fields on  $\mathcal P$ and $\mathcal V$.  In local coordinates $(Q^A,f^b)$,  they are    representedas as
 $K^A_{\alpha}(Q)\partial / \partial Q^A$
 with
$K^A_{\alpha}(Q)=\partial {\tilde Q}^A/\partial a^{\alpha}|_{a=e}$ and
$K^b_{\alpha}(f)\partial/\partial f^b$ with 
$$K^b_{\alpha}(f)=\partial {\tilde f}^b/\partial a^{\alpha}|_{a=e}=\partial {\bar D}^b_c(a)/\partial a^{\alpha}|_{a=e}f^c\equiv({\bar J}_{\alpha})^b_c f^c.$$  
The generators ${\bar J}_{\alpha}$ of the representation ${\bar D}^b_c(a)$ satisfy the  commutation relation 
$[{\bar J}_{\alpha},{\bar J}_{\beta}]={\bar c}^{\gamma}_{\alpha \beta}{\bar J}_{\gamma}$, where the structure constants
${\bar c}^{\gamma}_{\alpha \beta}=-{c}^{\gamma}_{\alpha \beta}$.


Based on \cite{AbrMarsd}, we conclude that the  action of a group $\mathcal G$ on the manifold  $\tilde{\mathcal P}$  leads to the  principal fiber bundle $ \pi':\mathcal P\times \mathcal V \to \mathcal P\times _{\mathcal G}\mathcal V$.\footnote{$\pi':(p,v)\to [p,v]$, where  $[p,v]$ is the equivalence class formed  by the  relation   $(p,v)\sim (pg,g^{-1}v).$} This allow us to regard the initial manifold $\tilde{\mathcal P}$ as a total space of this principal fiber bundle $\rm P(\tilde{\mathcal M},\mathcal G)$. By $\tilde{\mathcal M}$ we denoted the orbit space manifold $\mathcal P\times _{\mathcal G}\mathcal V$ -- the base space of the bundle $ \pi'$.

From this it follows that we can express the local coordinates $(Q^A, f^n)$ of the point $(p,v)\in \tilde{\mathcal P}$ in terms of the  coordinates of the principal fiber bundle. It is done with the help of the adapted coordinates. They are the group coordinates $a^{\alpha}(Q)$ and the invariant coordinates  $x^i(Q)$ determined from the equation
 \[
 Q^{\ast A}(x^i)=F^A(Q,a^{-1}(Q))),
\]
which satisfy $\chi^{\alpha}(Q^{\ast A}(x^i))=0$, where $\chi^{\alpha}(Q)$ is a ``gauge''- the function by which the local surface 
$\Sigma$ in $\mathcal P$ is defined.  This surface is needed to determine the coordinates in the principal fiber bundle.

The group coordinates $a^{\alpha}(Q)$ of a point $p\in \mathcal P$ are defined by the solution of the following equation:
\[
 \chi^{\alpha}(F^A(Q,a^{-1}(Q)))=0.
\]
This group element  carries the point $p$ to the submanifold  $\Sigma$, so that

 $Q^{\ast A}=F^A(Q,a^{-1}(Q)))$.

 With the principal fiber bundle coordinates, the point $(p,v)\in \tilde{\mathcal P}$, whose  coordinates were $(Q^A,f^b)$, obtains the adapted coordinates  $(x^i(Q),\tilde f^a(f), a^{\alpha}(Q))$. The replacement of the coordinates $(Q^A,f^b)$ of a point $(p,v)$ for a new coordinates is performed as follows:
\begin{equation}
Q^A=F^A(Q^{\ast}(x^i),a^{\alpha}),\;\;\;f^b=\bar D^b_c(a)\tilde f^c,
\label{transf_coord}
\end{equation}
where $a^{\alpha}=a^{\alpha}(Q)$ is obtained as before.
 
In a new coordinates the  Riemannian metric can be represented in the following form: 
\begin{equation}
\displaystyle
G_{\tilde A \tilde B}=
\left(
\begin{array}{ccc}
{\tilde h}_{ij}+{\mathscr A}_i^\mu {\mathscr A}_j^\nu d_{\mu \nu } & 0 & {\mathscr A}_i^\mu
d_{\mu\nu}
\bar{u}^{\nu}_{\alpha}(a) \\ 
0 & G_{ab}  & {\mathscr A}^{\mu}_a d_{\mu\nu}{\bar u}^{\nu}_{\alpha}(a)\\
{\mathscr A}^{\mu}_jd_{\mu\nu}{\bar u}^{\nu}_{\beta}(a) & {\mathscr A}^{\mu}_b d_{\mu\nu}{\bar u}^{\nu}_{\beta}(a) & d_{\mu \nu}{\bar u}^{\mu}_{\alpha}(a){\bar u}^{\nu}_{\beta}(a)\\
\end{array}
\right),
\label{transfmetric}
\end{equation}
where 
${\tilde h}_{ij}(x,\tilde f)=Q^{\ast}{}^A_i{\tilde G}^{\rm H}_{AB}Q^{\ast}{}^B_j$, and ${\tilde G}^{\rm H}_{AB}=G_{AB}-G_{AC}K^C_{\mu}d^{\mu\nu}K^D_{\nu}G_{DB}.$
Further, we will also denote expressions that include $d^{\mu\nu}$, with a tilde mark above the character associated with that expression.

 The metric is written in terms of the components $({\mathscr A}^{\alpha}_i, {\mathscr A}^{\alpha}_p)$ of the mechanical connection  that exists in the principal fiber bundle $\rm P(\tilde{\mathcal M},\mathcal G)$. These components are determined  as follows:
$${\mathscr A}^{\alpha}_i(x,\tilde f)=d^{\alpha \beta}K^C_{\beta}G_{DC}Q^{\star}{}^D_i, \;\;\;{\mathscr A}^{\alpha}_p(x,\tilde f)=d^{\alpha \beta}K^r_{\beta}G_{rp}.$$

The inverse matrix $G^{\tilde A \tilde B}$  is given by
\begin{equation}
 \displaystyle
G^{\tilde A \tilde B}=\left(
\begin{array}{ccc}
 h^{ij} & \underset{\scriptscriptstyle{(\gamma)}}{{\mathscr A}^{\mu}_m} K^a_{\mu} h^{mj} & -h^{nj}\,\underset{\scriptscriptstyle{(\gamma)}}{{\mathscr A}^{\beta}_n} \bar v ^{\alpha}_{\beta} \\
\underset{\scriptscriptstyle{(\gamma)}}{{\mathscr A}^{\mu}_m} K^b_{\mu} h^{ni} & G^{AB}N^a_AN^b_B+G^{ab} & -G^{EC}{\Lambda}^{\beta}_E{\Lambda}^{\mu}_CK^b_{\mu}\bar v ^{\alpha}_{\beta}
\\
-h^{ki}\underset{\scriptscriptstyle{(\gamma)}}{{\mathscr A}^{\varepsilon}_k}\bar v ^{\beta}_{\varepsilon} & -G^{EC}{\Lambda}^{\varepsilon}_E{\Lambda}^{\mu}_CK^a_{\mu}\bar v ^{\beta}_{\varepsilon} & G^{BC}{\Lambda}^{\alpha'}_B{\Lambda}^{\beta'}_C\bar v ^{\alpha}_{\alpha'}v ^{\beta}_{\beta'}
\end{array}
\right).
\label{invers_metric}
\end{equation}
Here ${\Lambda}^{\beta}_E=(\Phi^{-1})^{\beta}_{\mu}{\chi}^{\mu}_E$,  $h^{ij}$ is an inverse matrix to the matrix $h_{ij}=Q^{\ast A}_i G^{\rm H}_{AB}Q^{\ast B}_j$ with $$G^{\rm H}_{AB}=G_{AB}-G_{AD}K^D_{\alpha}{\gamma}^{\alpha\beta}K^C_{\beta}G_{CB}.$$ 
By $\underset{\scriptscriptstyle{(\gamma)}}{{\mathscr A}^{\mu}_m}$ we denote the mechanical connection formed from the orbit metric $\gamma_{\mu \nu}$:
$$\underset{\scriptscriptstyle{(\gamma)}}{{\mathscr A}^{\mu}_m}={\gamma}^{\mu \nu}K^A_{\nu}G_{AB}Q^{\ast B}_m.$$
By $\chi^{\alpha}_B$ we denote $\chi^{\alpha}_B=\partial \chi^{\alpha}(Q)/\partial Q^B |_{Q=Q^{\ast}(x)}$, $(\Phi)^{\alpha}_{\beta}=K^A_{\beta}\chi^{\alpha}_A$ is the Faddeev-Popov matrix, $N^b_B=-K^b_{\mu}(\Phi)^{\mu}_{\nu}{\chi}^{\nu}_B \equiv -K^b_{\mu}{\Lambda}^{\mu}_B$ is one of the components of a particular projector on a tangent space to the orbit space.

The determinant of matrix (\ref{transfmetric}) is equal to
\begin{eqnarray}
 \det G_{\tilde A \tilde B}=(\det d_{\alpha\beta})\,(\det {\bar u}^{\mu}_{\nu}(a))^2 \displaystyle
\det \left(
\begin{array}{cc}
\tilde h_{ij} & \tilde G^{\rm H}_{B b}Q_{i}^{*B}\\
\tilde G^{\rm H}_{Aa}Q_{j}^{*A} & \tilde G^{\rm H}_{ba}\\
\end{array}
\right),
\label{det}
\end{eqnarray}
where $\tilde G^{\rm H}_{Aa}=-G_{AB}K^B_{\mu}d^{\mu\nu}K^b_{\nu}G_{ba}$, $\tilde G^{\rm H}_{ba}=G_{ba}-G_{bc}K^c_{\mu}d^{\mu\nu}K_{\nu}^pG_{pa}$.

Note that the last determinant on the right hand side of (\ref{det}) is the determinant of the metric  defined on the orbit space $\tilde{\mathcal M}=
\mathcal P\times_{\mathcal G}\mathcal V$ of the principal fiber bundle $\rm P(\tilde{\mathcal M},\mathcal G).$
In what follows we will denote this determinant by $H$.

Also note that in the matrix $G^{\tilde A \tilde B}$,  the upper left quadrant of the matrix 
(\ref{invers_metric}) is ​​the matrix that represents the inverse metric to the metric in the orbit space of our principal fiber bundle.

The introduction of new coordinates leads to a transformation of the original stochastic process, the measures that they generate, and, consequently, the local evolutionary semigroups. By gluing together these local semigroups on their common intersection domains, we obtain, after passing to the corresponding limit, a new global semigroup. 

The next local semigroup transformation performed in \cite{Storchak_2019} was a transformation that leads to factorization of path integral measures in local semigroups, which allows obtaining the corresponding global semigroup with factorized measure.

 Thus, as a result of factorization of the measure in the original path integral (\ref{orig_path_int})
           it became possible to derive  
 the integral relation between the Green functions given on the total space $\tilde{\mathcal P}$ of the principal fiber bundle $\rm P(\tilde {\mathcal M},\mathcal G)$ and on the orbit space $\tilde{\mathcal M}$ of this bundle. 
The obtained integral relation looks as follows:
\begin{eqnarray}
&&G^{\lambda}_{mn}(x_b,\tilde f_b,t_b;
 x_a,\tilde f_a,t_a)=
\displaystyle\int _{\cal G}G_{\tilde{\cal P}}(p_b\theta,v_b\theta,t_b;
p_a,t_a) 
D_{nm}^\lambda (\theta )d\mu (\theta ),
\nonumber\\
&&(x,\tilde f)=\pi'(p,v).
\label{green_funk_relat}
\end{eqnarray}

The path integral which represent the Green function $G^{\lambda}_{mn}$ is written symbolically as
\begin{eqnarray}
&&G^{\lambda}_{mn}(\pi'(p_b),\pi'(v_b),t_b;
 \pi'(p_a),\pi'(v_a),t_a)=
\nonumber\\
&&=\int\limits_{{\xi (t_a)=\pi' (p_a,v_a)}\atop  
{\xi (t_b)=\pi' (p_b,v_b)}} d{\mu}^{\xi}
\exp \{\frac 1{\mu ^2\kappa
m}\int_{t_a}^{t_b}\tilde{V}(\xi_1(u),\xi_2(u))du\}
\nonumber\\
&&\times
\overleftarrow{\exp }%
\int_{t_a}^{t_b}\Bigl\{{\mu}^2\kappa\Bigl[\frac 12d^{\alpha \nu
}(x(u),\tilde f(u))(J_\alpha
)_{pr}^\lambda (J_\nu )_{rn}^\lambda \Bigr.\Bigr.
\nonumber\\
&&-\Bigl.\Bigl.\frac 12\frac 1{\sqrt{d\, H}}\frac \partial {\partial x^k}\left( \sqrt{d\, H}%
h^{km}\underset{\scriptscriptstyle{(\gamma)}}{{\mathscr A}_m^\nu}(x(u)) \right) (J_\nu )_{pn}^\lambda  
\nonumber\\
&&-\frac12(G^{EC}\Lambda^{\nu}_E\Lambda^{\mu}_C)\frac 1{\sqrt{d\, H}}\frac \partial {\partial \tilde f^b}\left(\sqrt{d\, H}K^b_{\mu}\right)(J_\nu )_{pn}^\lambda \Bigr]du
\nonumber\\
&&-\mu\sqrt{\kappa}\Bigl[\underset{\scriptscriptstyle{(\gamma)}}{{\mathscr A}^{\nu}_k}(x(u))\tilde X^k_{\bar m}(u)(J_\nu )_{pn}^\lambda d\tilde w^{\bar m}(u)+\tilde {\mathscr A}^{\nu}_a\tilde X^a_{\bar b}(u)(J_\nu )_{pn}^\lambda d\tilde w^{\bar b}(u)\Bigr]\Bigr\}.
\label{path_int_G_mn}
\end{eqnarray}

The differential generator  of  the matrix semigroups with the kernel $G^{\lambda}_{mn}$ (without the potential term) is given by 
\begin{eqnarray}
&&\frac 12\mu ^2\kappa \Bigl\{\Bigl[\triangle _{\tilde M}
+h^{ni}\Bigl(\frac 1{\sqrt{d}%
}\frac {\partial (\sqrt{d}\,)}
{\partial x^n}+\underset{\scriptscriptstyle{(\gamma)}}{{\mathscr A}^{\nu}_n}\frac 1{\sqrt{d}%
}\frac {\partial (\sqrt{d}K^b_{\nu}\,)}
{\partial \tilde f^b}\Bigr)\frac \partial
{\partial x^i}
\nonumber\\
&&+\Bigl(h^{mi}\underset{\scriptscriptstyle{(\gamma)}}{{\mathscr A}^{\mu}_m}K^a_{\mu}\frac 1{\sqrt{d}%
}\frac {\partial (\sqrt{d}\,)}{\partial  x^i}+(G^{ab}+G^{AB}N^a_AN^b_B)\frac 1{\sqrt{d}%
}\frac {\partial (\sqrt{d}\,)}
{\partial \tilde f^b}\Bigr)\frac \partial
{\partial \tilde f^a}
\Bigr](I^\lambda )_{pq}
\nonumber\\
&&-2h^{ni}\underset{\scriptscriptstyle{(\gamma)}}{{\mathscr A}^{\beta}_n} (J_\beta )_{pq}^\lambda
\frac \partial {\partial x^i}-2h^{nk}\underset{\scriptscriptstyle{(\gamma)}}{{\mathscr A}^{\beta}_n} \underset{\scriptscriptstyle{(\gamma)}}{{\mathscr A}^{\mu}_k}K^a_{\mu}(J_\beta )_{pq}^\lambda \frac \partial {\partial \tilde f^a}
\bigr.
\nonumber\\
&&-2({\gamma}^{\alpha\beta}K^a_{\alpha}K^b_{\beta}+G^{ab})\tilde {\mathscr A}^{\beta}_a(J_\beta )_{pq}^\lambda \frac \partial {\partial \tilde f^b}
\bigr.
\nonumber\\
&&-\bigl.\Bigl[\frac 1{\sqrt{dH}}\frac
\partial {\partial x^i}\left( \sqrt{dH}h^{ni}\underset{\scriptscriptstyle{(\gamma)}}{{\mathscr A}^{\beta}_n}\right)  
+(G^{EC}\Lambda^{\beta}_E\Lambda^{\mu}_C)\frac 1{\sqrt{dH}}
\frac {\partial (\sqrt{dH}K^b_{\mu}\,)}{\partial  \tilde f^b}\Bigr](J_\beta )_{pq}^\lambda 
\nonumber\\
&&+({\gamma }^{\alpha \beta }+h^{ij}\underset{\scriptscriptstyle{(\gamma)}}{{\mathscr A}^{\alpha}_i}\underset{\scriptscriptstyle{(\gamma)}}{{\mathscr A}^{\beta}_j})(J_\alpha
)_{pq^{\prime }}^\lambda (J_\beta )_{q^{\prime }q}^\lambda \Bigr\}.
\label{dif_gen_ksi}
\end{eqnarray} 
(Here  $(I^\lambda )_{pq}$ is a unity matrix.) 

This operator  acts in the space of the sections 
$\Gamma ({\tilde{\cal M}},V^{*})$ of the associated co-vector 
bundle 
${\cal E}^{*}={\tilde{\cal P}}\times _{\cal G}V^{\ast}_\lambda $ ($ {\tilde{\cal P}}=\cal P\times\cal \mathcal V$) with 
the scalar product  
\begin{equation*}
(\psi _n,\psi _m)=\int_{\tilde{\cal M}}\langle \psi _n,\psi _m{\rangle}_
{V^{\ast}_{\lambda}}
 \sqrt{d(x,\tilde f)}dv_{\tilde{\cal M}}(x,\tilde f),
\end{equation*}
 where $dv_{\tilde{\cal M}}(x,\tilde f)=\sqrt{H(x,\tilde f)}dx^1...dx^{n_{\cal M}}d\tilde f^1...\tilde f^{n_{\cal \mathcal V}}$ 
is the Riemannian volume element on the manifold ${\tilde{\cal M}}$. 

The measure ${\mu}^{{\xi}}$ in (\ref{path_int_G_mn}) is generated by  
the stochastic process $\xi(t)$ given by  the solution of the stochastic differential equation on the manifold $\tilde{\mathcal M}$. The local form of this equation is represented by an equation that combines the equations for the local processes $ x^i(t)$ and $ \tilde f^a (t)) $: 
\begin{equation}
 d\xi^i_{\rm loc}(t)=\frac12\mu^2\kappa{b^i\choose b^a}dt+\mu\sqrt{\kappa}{\tilde X^i_{\bar m}\;\; 0\choose {\tilde X^a_{\bar m}\;\; \tilde X^a_{\bar b}}} {d\tilde w^{\bar m}\choose d\tilde w^{\bar b}}\;\;\;
\label{sde-chi}
\end{equation}
with
\[
 {b^i\choose b^a}={\tilde b^i\choose \tilde b^a}+{h^{ij} \;\: \;\;\;\;\;\;\;\;\;\underset{\scriptscriptstyle{(\gamma)}}{{\mathscr A}^{\mu}_m} K^b_{\mu} h^{mi} 
\choose { \underset{\scriptscriptstyle{(\gamma)}}{{\mathscr A}^{\mu}_m} K^a_{\mu} h^{nj} \;\; G^{AB}N^a_AN^b_B+G^{ab}}}{\frac{1}{\sqrt{d}}\frac{\partial}{\partial x^j}\sqrt{d}\choose \frac{1}{\sqrt{d}}\frac{\partial}{\partial \tilde f^b}\,\sqrt{d} \; },
\]
\[
{\rm in}\;\;{\rm which}\;\; \;\;\;\;\;{h^{ij} \;\; \;\;\;\;\;\;\;\;\;\underset{\scriptscriptstyle{(\gamma)}}{{\mathscr A}^{\mu}_m} K^b_{\mu} h^{mi} 
\choose { \underset{\scriptscriptstyle{(\gamma)}}{{\mathscr A}^{\mu}_m} K^a_{\mu} h^{nj} \;\; G^{AB}N^a_AN^b_B+G^{ab}}}={{\tilde h}^{ij} \;\; \;{\tilde h}^{ib} 
\choose { {\tilde h}^{aj} \;\; {\tilde h}^{ab}}},
\]
\[
 \tilde b^i=\frac1{\sqrt{H}}\frac{\partial}{\partial x^j}\Bigl(\sqrt{H}h^{ij}\Bigr)+\underset{\scriptscriptstyle{(\gamma)}}{{\mathscr A}^{\mu}_n}h^{ni}\frac1{\sqrt{H}}\frac{\partial}{\partial \tilde f^b}\Bigl(\sqrt{H}K^b_{\mu}\Bigr)\;\;\;\rm{and}
\]
\begin{eqnarray*}
 \tilde b^a&=&\frac1{\sqrt{H}}\frac{\partial}{\partial x^j}\Bigl(\sqrt{H}h^{mj}\underset{\scriptscriptstyle{(\gamma)}}{{\mathscr A}^{\mu}_m}\Bigr)K^a_{\mu}+(G^{ab}+G^{AB}N^a_AN^b_B)\frac1{\sqrt{H}}\frac{\partial}{\partial \tilde f^b}\Bigl(\sqrt{H}\Bigr)
\\
&&+\frac{\partial}{\partial \tilde f^b}\Bigl(G^{AB}N^a_AN^b_B\Bigr).
\end{eqnarray*}

\section{Girsanov transformation of the path integral measure}
The drift coefficients $b^i$ and $b^a$ of the equation (\ref{sde-chi}) include the  terms with the  partial derivatives of  the determinant   of the metric $d_{\alpha\beta}(x,\tilde f)$  given on the orbits of the principal fiber bundle $\rm P(\tilde {\mathcal M},\mathcal G)$.  

These terms are not intrinsic for the orbit space $\tilde{\mathcal M}$ and hence are not necessary for description of the evolution on this space.
Therefore, we must perform such a transformation of the path integral in which the measure $ {\mu}^{{\xi}} $ is replaced by a new measure $ {\mu}^{\tilde {\xi}} $ associated with a new stochastic process $ \tilde \xi (t) $ with the following local stochastic differential equations:
\begin{equation}
 d\tilde\xi^i_{\rm loc}(t)=\frac12\mu^2\kappa{\tilde b^i\choose \tilde b^a}dt+\mu\sqrt{\kappa}{\tilde X^i_{\bar m}\;\; 0\choose {\tilde X^a_{\bar m}\;\; \tilde X^a_{\bar b}}} {d\tilde w^{\bar m}\choose d\tilde w^{\bar b}}.
\label{sde-tilde-chi}
\end{equation}
Such transformation is known as the Girsanov transformation of the stochasic processes.
This transformation is based on the assumption that the partial differential equation whose solution is represented by the path integral have a unique solution.
In our case, this allows us to find a rule according to which the multiplicative stochastic integral under the path integral sign in 
(\ref{path_int_G_mn}) is transformed. 

The Girsanov transformation means that the following equality of the path integrals must hold:
\begin{equation}
 \int \;d{\mu}^{\xi}\; \overleftarrow{\exp }_{\xi}(...)^{\lambda}_{pq}{\varphi}_p(\xi)=\int \;d{\mu}^{\tilde{\xi}}\; \overleftarrow{\exp }_{\tilde{\xi}}(...)^{\lambda}_{pq}{\varphi}_p(\tilde\xi),\footnote{Here the summation over repeated indices is implied.}
\label{semigroup}
\end{equation}
where the multiplicative stochastic integral $\overleftarrow{\exp }_{\xi}(...)^{\lambda}_{pq}$ is as in (\ref{path_int_G_mn}). As for the multiplicative stochastic integral $\overleftarrow{\exp }_{\tilde{\xi}}(...)^{\lambda}_{pq}$, we assume that it has the following form: 
\[
 \overleftarrow{\exp }_{\tilde{\xi}}(...)^{\lambda}_{pq}=\Bigl(\overleftarrow{\exp }_{\tilde{\xi}}\int ({\tilde L}dw+{\tilde M}'dt)\Bigr)^{\lambda}_{pq},
\]
where 
\[
 ``{\tilde L}dw{}\textquotedblright=-(\mu\sqrt{\kappa})\Bigl((\tilde{\Gamma}_{3n})^{\lambda}_{pq}\tilde{X}^n_{\bar m}d{\tilde w}^{\bar m}+(\tilde{\Gamma}_{4c})^{\lambda}_{pq}\tilde{X}^c_{\bar b}d{\tilde w}^{\bar b}\Bigr),
\]
and
\[
 ``{\tilde M'}\textquotedblright=(\mu ^2 \kappa){\tilde M'}{}^{\lambda}_{pq}dt
\]

\[
  \overleftarrow{\exp }_{\tilde{\xi}}(...)=\overleftarrow{\exp }_{\tilde{\xi}}\int \biggl((\mu ^2 \kappa){\tilde M'}{}^{\lambda}_{pq}dt-(\mu\sqrt{\kappa})\Bigl((\tilde{\Gamma}_{3n})^{\lambda}_{pq}\tilde{X}^n_{\bar m}d{\tilde w}^{\bar m}+((\tilde{\Gamma}_{4c})^{\lambda}_{pq}\tilde{X}^c_{\bar b}d{\tilde w}^{\bar b}\Bigr)\biggr).
\]
The  unknown coefficients $\tilde{\Gamma}_{3n}$, $\tilde{\Gamma}_{4c}$ and ${\tilde M}'$ can be determined by comparing the differential generators   for the  evolution matrix semigroups that represented by the path integrals in (\ref{semigroup}). 

The differential generator for the matrix semigroup defined by the left-hand side of (\ref{semigroup}) is (\ref{dif_gen_ksi}). The differential generator for the matrix semigroup, given on the right-hand side of (\ref{semigroup}), can be obtained by taking the mathematical expectation of the Ito's differential of $\varphi _p(\tilde\xi)\tilde Z^{\lambda}_{pq}(\tilde\xi)$:
\[
 d\bigl(\varphi _p\,\tilde Z^{\lambda}_{pq}\bigr)=d\varphi _p\,Z^{\lambda}_{pq}+\varphi _p\,dZ^{\lambda}_{pq}+d\varphi _p\,d\tilde Z^{\lambda}_{pq},
\]
where  $\tilde Z^{\lambda}_{pq}$ is notation for $\overleftarrow{\exp }_{\tilde{\xi}}(...)^{\lambda}_{pq}$.

 Taking the mathematical expectation  of the terms represented by  $d\varphi _p\,Z^{\lambda}_{pq}$, we get the following expression:
\begin{eqnarray*}
 &&\Bigl(\frac{\partial \varphi _p}{\partial x^i}{\tilde b}^i+\frac{\partial \varphi _p}{\partial \tilde f^a}{\tilde b}^a+\frac12\frac{\partial ^2\varphi _p}{\partial x^i\partial x^j}\tilde X^i_{\bar m}\tilde X^j_{\bar m}+\frac12\frac{\partial ^2\varphi _p}{\partial \tilde f^a\partial \tilde f^b}(\tilde X^a_{\bar m}\tilde X^b_{\bar m}+\tilde X^a_{\bar a}\tilde X^b_{\bar a})
\nonumber\\
&&\; +\frac{\partial ^2\varphi _p}{\partial x^i\partial\tilde f^a}(\tilde X^i_{\bar m}\tilde X^a_{\bar m}+\tilde X^i_{\bar b}\tilde X^a_{\bar b})\Bigr)({\rm I}^{\lambda})_{pq}.
\end{eqnarray*}
This expression can  also be represented as follows:
\begin{equation}
 \Bigl(\frac{\partial \varphi _p}{\partial x^i}\,{\tilde b}^i+\frac{\partial \varphi _p}{\partial \tilde f^a}\,{\tilde b}^a+\frac12\frac{\partial ^2\varphi _p}{\partial x^i\partial x^j}\,\tilde h^{ij} +\frac12\frac{\partial ^2\varphi _p}{\partial \tilde f^a\partial \tilde f^b}\,\tilde h^{ab}
 +\frac{\partial ^2\varphi _p}{\partial x^i\partial\tilde f^a}\,\tilde h^{ia}\Bigr)({\rm I}^{\lambda})_{pq}.
\label{partial_x_i}
\end{equation}

The mathematical expectation  of the terms given by $d\varphi _p\,d\tilde Z^{\lambda}_{pq}$ are
\begin{eqnarray*}
 -\Bigl(\frac{\partial \varphi _p}{\partial x^i}\Bigr)\tilde X^i_{\bar m}(\tilde{\Gamma}_{3n})^{\lambda}_{pq}\tilde{X}^n_{\bar m}-\Bigl(\frac{\partial \varphi _p}{\partial \tilde f^a}\Bigr)\Bigl((\tilde{\Gamma}_{3n})^{\lambda}_{pq}\tilde{X}^n_{\bar m}\tilde X^a_{\bar m}+(\tilde{\Gamma}_{4c})^{\lambda}_{pq}\tilde{X}^c_{\bar b}\tilde X^a_{\bar b}\Bigr),
\end{eqnarray*}
or written in another form
\begin{equation}
  -\Bigl(\frac{\partial \varphi _p}{\partial x^i}\Bigr)\tilde h^{in}(\tilde{\Gamma}_{3n})^{\lambda}_{pq}-\Bigl(\frac{\partial \varphi _p}{\partial \tilde f^a}\Bigr)\Bigl((\tilde{\Gamma}_{3n})^{\lambda}_{pq}\tilde{h}^{na}+(\tilde{\Gamma}_{4c})^{\lambda}_{pq}\tilde{h}^{ac}\Bigr), 
\label{partial_f_a}
\end{equation}
where $\tilde h^{in}=h^{in}$, $\tilde h^{an}=\underset{\scriptscriptstyle{(\gamma)}}{{\mathscr A}^{\mu}_m}K^a_{\mu}h^{mn}$, $\tilde h^{ac}=(\gamma^{\alpha\beta}K^a_{\alpha}K^c_{\beta}+G^{ac}).$

Comparing those terms of the differential operators that stand at the first derivatives of $\varphi _p$ in (\ref{dif_gen_ksi}) with the analogous terms  in (\ref{partial_x_i}) and (\ref{partial_f_a}), one  can  obtain that 
\[
 (\tilde{\Gamma}_{3n})^{\lambda}_{pq}=\underset{\scriptscriptstyle{(\gamma)}}{{\mathscr A}^{\beta}_n}(J_\beta )_{pq}^\lambda-\frac12\frac{1}{\sqrt{d}}\frac{\partial (\sqrt{d})}{\partial x^n}({\rm{I}^{\lambda}})_{pq}-\frac12\underset{\scriptscriptstyle{(\gamma)}}{{\mathscr A}^{\nu}_n}K^b_{\nu}\frac{1}{\sqrt{d}}\frac{\partial (\sqrt{d})}{\partial \tilde f^b}(\rm{I}^{\lambda})_{pq}
\]
and
\[
 (\tilde{\Gamma}_{4c})^{\lambda}_{pq}=-\frac12\frac{1}{\sqrt{d}}\frac{\partial (\sqrt{d})}{\partial \tilde f^c}({\rm{I}^{\lambda}})_{pq}+\tilde{\mathscr A}^{\beta}_c(J_\beta )_{pq}^\lambda.
\]

The linear in $\varphi _p$ terms in the differential generator  (\ref{dif_gen_ksi}), and      obtained   from $\varphi _p\,dZ^{\lambda}_{pq}$ terms in the differential generator for the matrix semigroup related with the process $\tilde{\xi}$, are used to define the drift term $\tilde M'{}^{\lambda}_{pq}$ in the  multiplicative stochastic integral.

  \textquotedblleft $\varphi _p\,dZ^{\lambda}_{pq}$''-- terms
 are given by
\begin{equation}
 (\mu ^2\kappa)\Bigl\{\tilde M'{}^{\lambda}_{pq}+\frac12\Bigl((\tilde{\Gamma}_{3n})^{\lambda}_{pq'}(\tilde{\Gamma}_{3k})^{\lambda}_{q'q}h^{nk}+(\tilde{\Gamma}_{4a})^{\lambda}_{pq'}(\tilde{\Gamma}_{4c})^{\lambda}_{q'q}\tilde h^{ac}\Bigr)\Bigr\}\,\varphi _p.
\label{fi_dZ}
\end{equation} 
$(\tilde{\Gamma}_{3n})^{\lambda}_{pq'}(\tilde{\Gamma}_{3k})^{\lambda}_{q'q}$ has the following explicit representation:
\begin{eqnarray*}
&&\underset{\scriptscriptstyle{(\gamma)}}{{\mathscr A}^{\beta}_n}\underset{\scriptscriptstyle{(\gamma)}}{{\mathscr A}^{\alpha}_k} (J_\beta )_{pq'}^\lambda(J_\alpha )_{q'q}^\lambda -\frac12 d_n\underset{\scriptscriptstyle{(\gamma)}}{{\mathscr A}^{\alpha}_k}(J_\alpha )_{pq}^\lambda -\frac12 d_bK^b_{\nu}\underset{\scriptscriptstyle{(\gamma)}}{{\mathscr A}^{\nu}_n}\underset{\scriptscriptstyle{(\gamma)}}{{\mathscr A}^{\alpha}_k}(J_\alpha )_{pq}^\lambda
\nonumber\\
&&-\frac12d_k\underset{\scriptscriptstyle{(\gamma)}}{{\mathscr A}^{\beta}_n}(J_\beta )_{pq}^\lambda
+\frac14d_nd_k({\rm I^{\lambda}})_{pq}+\frac14 d_bd_kK^b_{\nu}\underset{\scriptscriptstyle{(\gamma)}}{{\mathscr A}^{\nu}_n}\,({\rm I^{\lambda}})_{pq}
\nonumber\\
&&-\frac12 d_bK^b_{\alpha}\underset{\scriptscriptstyle{(\gamma)}}{{\mathscr A}^{\alpha}_k}\underset{\scriptscriptstyle{(\gamma)}}{{\mathscr A}^{\beta}_n}(J_\beta )_{pq}^\lambda+\frac14d_nd_bK^b_{\alpha}\underset{\scriptscriptstyle{(\gamma)}}{{\mathscr A}^{\alpha}_k}({\rm I^{\lambda}})_{pq}+\frac14d_bd_cK^b_{\nu}\underset{\scriptscriptstyle{(\gamma)}}{{\mathscr A}^{\nu}_k}K^c_{\alpha}\underset{\scriptscriptstyle{(\gamma)}}{{\mathscr A}^{\alpha}_k}({\rm I^{\lambda}})_{pq}.
\end{eqnarray*}
An explicit expression of $(\tilde{\Gamma}_{4a})^{\lambda}_{pq'}(\tilde{\Gamma}_{4c})^{\lambda}_{q'q}$ is 
\begin{equation*}
 \tilde{\mathscr A}^{\beta}_c\tilde{\mathscr A}^{\mu}_a(J_\beta )_{pq'}^\lambda(J_\mu )_{q'q}^\lambda -\frac12 d_c\tilde{\mathscr A}^{\mu}_a(J_\mu )_{pq}^\lambda -\frac12
d_a\tilde{\mathscr A}^{\beta}_c(J_\beta )_{pq}^\lambda +\frac14d_ad_c({\rm I^{\lambda}})_{pq}.
\end{equation*}
In the above formulas, it was used  a new  notation by which $d_n=\frac{1}{\sqrt{d}}\frac{\partial (\sqrt{d})}{\partial x^n}$ and $d_b=\frac{1}{\sqrt{d}}\frac{\partial (\sqrt{d})}{\partial \tilde f^b}$.

Thus, 
  $\tilde M'{}^{\lambda}_{pq}$ is defined from the equation  in which the left-hand side  is given by   (\ref{fi_dZ}), and the right-hand side of the equation consists of the corresponding terms obtained as a result of applying   the differential operator (\ref{dif_gen_ksi}) to $\varphi _p$.

First, it can be shown that in this equation there is a cancellation of the terms that include as a factor only one term  $d_i$ (or $d_a$) and $(J_\beta )_{pq}^\lambda$. 

For these terms we will have
\begin{eqnarray*}
&&d_bK^b_{\nu}h^{nk}\underset{\scriptscriptstyle{(\gamma)}}{{\mathscr A}^{\nu}_n}\underset{\scriptscriptstyle{(\gamma)}}{{\mathscr A}^{\alpha}_k} (J_\alpha )_{pq}^\lambda(-\frac14-\frac14)+d_a\tilde{\mathscr A}^{\beta}_c(J_\beta )_{pq}^\lambda (\gamma^{\alpha\beta}K^a_{\alpha}K^c_{\beta}+G^{ac})(-\frac14-\frac14)=
\nonumber\\
&&-\frac12d_bK^b_{\nu}(G^{EC}{\Lambda}^{\nu}_E{\Lambda}^{\mu}_C)(J_\mu )_{pq}^\lambda.
\end{eqnarray*}

Since $G^{EC}{\Lambda}^{\nu}_E{\Lambda}^{\mu}_C=\gamma^{\mu\nu}+h^{nk}\underset{\scriptscriptstyle{(\gamma)}}{{\mathscr A}^{\nu}_n}\underset{\scriptscriptstyle{(\gamma)}}{{\mathscr A}^{\alpha}_k}$,  we get
\[
 -d_ad^{\beta\sigma}K^e_{\sigma}G_{ec}(\gamma^{\mu\nu}K^a_{\mu}K^c_{\nu}+G^{ac})=-d_bK^b_{\nu}\gamma^{\beta\nu}.
\]
But this equality lead to the identity
$\;\;
 d_aK^a_{\mu}\bigl(d^{\beta\sigma}{\gamma'}_{\sigma\nu}\gamma^{\mu\nu}+d^{\beta\mu}-\gamma ^{\mu\beta}\equiv0\bigr).$\\
( We recall that $d_{\alpha\beta}=\gamma_{\alpha\beta}+\gamma'_{\alpha\beta}$ and $\gamma'_{\alpha\beta}=K^a_{\alpha}G_{ab}K^b_{\beta}$.)

From the equation for $\tilde M'{}^{\lambda}_{pq}$ it follows that its solution can be represented as the sum of  three group of terms: 
\[
 \tilde M'{}^{\lambda}_{pq}=(\mu^2\kappa)\Bigl(\{...\}({\rm I^{\lambda}})_{pq}+\{...\}^{\beta}(J_\beta )_{pq}^\lambda +\{...\}^{\alpha\beta} (J_\alpha)_{pq'}^\lambda(J_\beta )_{q'q}^\lambda\Bigr).
\]

The diagonal part of the solution, $\{...\}({\rm I^{\lambda}})_{pq}$, is
\begin{eqnarray*}
 &&-\frac12\Bigl(\frac14h^{nk}d_nd_k+\frac14d_bd_kK^b_{\nu}\underset{\scriptscriptstyle{(\gamma)}}{{\mathscr A}^{\nu}_n}h^{nk}+\frac14d_nd_bK^b_{\alpha}\underset{\scriptscriptstyle{(\gamma)}}{{\mathscr A}^{\alpha}_k}h^{nk}
\nonumber\\
&&\;\;\;\;\;\;+\frac14d_ad_ch^{nk}K^a_{\nu}K^c_{\mu}
\underset{\scriptscriptstyle{(\gamma)}}{{\mathscr A}^{\nu}_n}\underset{\scriptscriptstyle{(\gamma)}}{{\mathscr A}^{\mu}_k}+\frac14(\gamma^{\mu\nu}K^a_{\mu}K^c_{\nu}+G^{ac})d_ad_c\Bigr)({\rm I^{\lambda}})_{pq}.
\end{eqnarray*}
By denoting $\sigma=\ln d$ so that $d_n=\frac12{\sigma}_n$, we rewrite above terms as a quadratic form consisting of  the partial derivatives of $\sigma$:
\[
 -\frac18(\frac14h^{nk}\sigma _n\sigma _k+\frac12\sigma _b\sigma _k\tilde h^{bk}+\frac14\sigma _a\sigma _c\tilde h^{ac})({\rm I^{\lambda}})_{pq}=-\frac{1}{32}<\partial\sigma,\partial \sigma>_{\tilde{\mathcal M}}({\rm I^{\lambda}})_{pq}.
\]

The second group of terms 
in the solution for $\tilde M'{}^{\lambda}_{pq}$ is given by
\begin{eqnarray*}
&&\{...\}^{\beta}(J_\beta )_{pq}^\lambda =-\frac 12\frac 1{\sqrt{ H}}\frac \partial {\partial x^k}\left( \sqrt{ H}%
h^{km}\underset{\scriptscriptstyle{(\gamma)}}{{\mathscr A}_m^\beta}(x(u)) \right) (J_\beta )_{pq}^\lambda  
\nonumber\\
&&\;\;\;\;\;\;\;\;\;\;\;\;\;-\frac12(G^{EC}\Lambda^{\beta}_E\Lambda^{\mu}_C)\frac 1{\sqrt{ H}}\frac \partial {\partial \tilde f^b}\left(\sqrt{ H}K^b_{\mu}\right)(J_\beta )_{pq}^\lambda .
\end{eqnarray*}

The last group can be obtained by transforming the following equality
\[
 \{...\}^{\alpha\beta}+\frac12(\gamma^{\alpha\beta}K^a_{\alpha}K^c_{\beta}+G^{ac})=\frac12\gamma^{\alpha\beta}
\]
 into $\{...\}^{\alpha\beta}+\frac12(-d^{\alpha\beta})=0$. This is done  using an  explicit representation for the connection $\tilde{\mathscr A}^{\beta}_c=d^{\beta\mu}K^b_{\mu}G_{bc}$, and by substituting $\gamma' _{\beta\mu}=d_{\beta\mu}-\gamma_{\beta\mu}$.

As a result of the performed calculation the diffusion terms in the  multiplicative stochastic integral become as follows:
\[
-(\tilde{\Gamma}_{3n})^{\lambda}_{pq}\tilde{X}^n_{\bar m}d{\tilde w}^{\bar m}=-\underset{\scriptscriptstyle{(\gamma)}}{{\mathscr A}^{\beta}_n}(J_\beta )_{pq}^\lambda
\tilde{X}^n_{\bar m}d{\tilde w}^{\bar m}+\frac14\bigl(\sigma_n \tilde{X}^n_{\bar m}d{\tilde w}^{\bar m}+\sigma _b \tilde{X}^b_{\bar m}d{\tilde w}^{\bar m}\bigr)({\rm{I}^{\lambda}})_{pq},
\]
where $\tilde{X}^b_{\bar m}=\underset{\scriptscriptstyle{(\gamma)}}{{\mathscr A}^{\nu}_n}K^b_{\nu}
\tilde{X}^n_{\bar m}$, and
\[
 -(\tilde{\Gamma}_{4c})^{\lambda}_{pq}\tilde{X}^c_{\bar b}d{\tilde w}^{\bar b}=-\tilde{\mathscr A}^{\beta}_c(J_\beta )_{pq}^\lambda\tilde{X}^c_{\bar b}d{\tilde w}^{\bar b}+\frac14\sigma _c\tilde{X}^c_{\bar b}d{\tilde w}^{\bar b}({\rm{I}^{\lambda}})_{pq}.
\]
After substitution of the obtained  diffusion and drift terms into $\tilde\xi$ -- dependent  multiplicative stochastic integral from (\ref{semigroup}), it will contain, together with the off-diagonal terms, the terms of the diagonal matrices. The exponential  represented by  the diagonal matrix can be factor out of the  multiplicative stochastic integral and can be treated independently. Therefore, the multiplicative stochastic integral on the right-hand side  (\ref{semigroup}) can be represented as follows:
 \begin{eqnarray}
&&\exp\frac14\int_{t_a}^{t_b}\Bigl\{ \mu\sqrt{\kappa}\bigl[\sigma_n \tilde{X}^n_{\bar m}d{\tilde w}^{\bar m}+ \sigma_b(\tilde{X}^b_{\bar m}d{\tilde w}^{\bar m}+\tilde{X}^b_{\bar a}d{\tilde w}^{\bar a})\bigr]
\nonumber\\
&&-\frac{1}{8}{\mu}^2\kappa<\partial\sigma,\partial \sigma>_{\tilde{\mathcal M}}du\,\,\Bigr\}({\rm I^{\lambda}})_{pq}
\nonumber\\
&&\times
\overleftarrow{\exp }%
\int_{t_a}^{t_b}\Bigl\{{\mu}^2\kappa\Bigl[\frac 12d^{\alpha \nu
}(x(u),\tilde f(u))(J_\alpha
)_{pr}^\lambda (J_\nu )_{rn}^\lambda \Bigr.\Bigr.
\nonumber\\
&&-\Bigl.\Bigl.\frac 12\frac 1{\sqrt{ H}}\frac \partial {\partial x^k}\left( \sqrt{ H}%
h^{km}\underset{\scriptscriptstyle{(\gamma)}}{{\mathscr A}_m^\nu}(x(u)) \right) (J_\nu )_{pn}^\lambda  
\nonumber\\
&&-\frac12(G^{EC}\Lambda^{\nu}_E\Lambda^{\mu}_C)\frac 1{\sqrt{ H}}\frac \partial {\partial \tilde f^b}\left(\sqrt{ H}K^b_{\mu}\right)(J_\nu )_{pn}^\lambda \Bigr]du
\nonumber\\
&&-\mu\sqrt{\kappa}\Bigl[\underset{\scriptscriptstyle{(\gamma)}}{{\mathscr A}^{\nu}_k}(x(u))\tilde X^k_{\bar m}(u)(J_\nu )_{pn}^\lambda d\tilde w^{\bar m}(u)+\tilde {\mathscr A}^{\nu}_a\tilde X^a_{\bar b}(u)(J_\nu )_{pn}^\lambda d\tilde w^{\bar b}(u)\Bigr]\Bigr\}
\label{multiplic_exp}
\end{eqnarray}

In the same way as in case of the reduction onto the zero momentum level\cite{Storchak_2019} the exponential  with the stochastic integral in the first factor  of the multiplicative stochastic integral (\ref{multiplic_exp}) can be transformed by the Ito's identity. As a result, we arrive at the following representation of the   first factor in (\ref{multiplic_exp}): 
\begin{eqnarray}
\left(\frac{exp(\sigma(x(t_b),\tilde f(t_b))}{exp(\sigma(x(t_a),\tilde f(t_a))}\right)^{1/4}
\exp\Bigl\{-\frac18\mu^2\kappa\int^{t_b}_{t_a}\bigl(\triangle_{\tilde{\cal M}}\sigma +\frac14<\partial\sigma,\partial \sigma>_{\tilde{\cal M}}\bigr)du\Bigr\},
\label{girs_result}
\end{eqnarray}
that is, to the path integral reduction Jacobian.

In \cite{Storchak_2020}, it was obtained  the geometrical representation of this Jacobian. The integrand $\tilde J$ of this Jacobiana was expressed as follows:  
\[
 \tilde J={\rm R}_{\tilde{\mathcal P}}-{\rm R}_{\tilde{\mathcal M}}-{\rm R}_{\mathcal G}-\frac14 d_{\mu\nu}{\mathscr F}^{\mu}_{\scriptscriptstyle A'\scriptscriptstyle B'}{\mathscr F}^{\nu \scriptscriptstyle A'\scriptscriptstyle B'}-||j||^2,
\]
where ${\rm R}_{\tilde {\mathcal P}}$ is the scalar curvature of the original manifold, 
${\rm R}_{\tilde {\mathcal M}}$ is the scalar curvature of the orbit space manifold  
$\tilde {\mathcal M}$, ${\rm R}_{\mathrm {\cal G}}$ is the scalar curvature of the orbit, ${\mathscr F}^{\mu}_{\scriptscriptstyle A'\scriptscriptstyle B'}$ is the curvature of the mechanical connection, $ ||j||^2$ is the trace of the square of the fundamental form $ j_{\alpha \beta}$ of the orbit taken on $\tilde{\mathcal M}$.

Thus, the Girsanov transformation allows us to rewrite   
the integral relation (\ref{green_funk_relat}) as follows:
\begin{equation}
d_b^{-1/4}d_a^{-1/4}
{\tilde G}^{\lambda}_{pq}(x_b,\tilde f_b, t_b;x_a,\tilde f_a,t_a)=\int_{\mathcal G}{G}_{\tilde{\mathcal P}}(p_b\theta,v_b\theta,t_b;p_a,v_a,t_a)D_{qp}^\lambda (\theta )
d\mu (\theta ),
\nonumber\\
\end{equation}
where $d_b=d(x_b,\tilde f_b)$, $d_a=d(x_a,\tilde f_a)$ and
 the  Green's function ${\tilde G}^{\lambda}_{pq}$ 
is given by the following path integral
\begin{eqnarray}
&&{\tilde G}^{\lambda}_{mn}(\pi'(p_b),\pi'(v_b),t_b;
 \pi'(p_a),\pi'(v_a),t_a)=
\nonumber\\
&&=\int\limits_{{\tilde\xi (t_a)=\pi' (p_a,v_a)}\atop  
{\tilde\xi (t_b)=\pi' (p_b,v_b)}} d{\mu}^{\tilde\xi}
\exp \{\int_{t_a}^{t_b}\Bigl(\frac{\tilde{V}({\tilde \xi} (u))}{\mu ^2\kappa
m}-\frac18\mu ^2\kappa
{\tilde J}\Bigr)
du\biggr\}
\nonumber\\
&&\times
\overleftarrow{\exp }%
\int_{t_a}^{t_b}\Bigl\{{\mu}^2\kappa\Bigl[\frac 12d^{\alpha \nu
}(x(u),\tilde f(u))(J_\alpha
)_{pr}^\lambda (J_\nu )_{rn}^\lambda \Bigr.\Bigr.
\nonumber\\
&&-\Bigl.\Bigl.\frac 12\frac 1{\sqrt{ H}}\frac \partial {\partial x^k}\left( \sqrt{ H}%
h^{km}\underset{\scriptscriptstyle{(\gamma)}}{{\mathscr A}_m^\nu}(x(u)) \right) (J_\nu )_{pn}^\lambda  
\nonumber\\
&&-\frac12(G^{EC}\Lambda^{\nu}_E\Lambda^{\mu}_C)\frac 1{\sqrt{ H}}\frac \partial {\partial \tilde f^b}\left(\sqrt{ H}K^b_{\mu}\right)(J_\nu )_{pn}^\lambda \Bigr]du
\nonumber\\
&&-\mu\sqrt{\kappa}\Bigl[\underset{\scriptscriptstyle{(\gamma)}}{{\mathscr A}^{\nu}_k}(x(u))\tilde X^k_{\bar m}(u)(J_\nu )_{pn}^\lambda d\tilde w^{\bar m}(u)+\tilde {\mathscr A}^{\nu}_a\tilde X^a_{\bar b}(u)(J_\nu )_{pn}^\lambda d\tilde w^{\bar b}(u)\Bigr]\Bigr\}.
\label{path_int_G_mn_res}
\end{eqnarray}

The differential generator  of  the matrix semigroups with the kernel ${\tilde G}^{\lambda}_{mn}$  is given by 
\begin{eqnarray}
&&\frac 12\mu ^2\kappa \Bigl\{\Bigl[\triangle _{\tilde M}
+\frac{2 {\tilde V}}{(\mu ^2\kappa )^2 m}-\frac14{\tilde J}\Bigr](I^\lambda )_{pq}
\nonumber\\
&&-2h^{ni}\underset{\scriptscriptstyle{(\gamma)}}{{\mathscr A}^{\beta}_n} (J_\beta )_{pq}^\lambda
\frac \partial {\partial x^i}-2h^{nk}\underset{\scriptscriptstyle{(\gamma)}}{{\mathscr A}^{\beta}_n} \underset{\scriptscriptstyle{(\gamma)}}{{\mathscr A}^{\mu}_k}K^a_{\mu}(J_\beta )_{pq}^\lambda \frac \partial {\partial \tilde f^a}
\bigr.
\nonumber\\
&&-2({\gamma}^{\alpha\beta}K^a_{\alpha}K^b_{\beta}+G^{ab})\tilde {\mathscr A}^{\beta}_a(J_\beta )_{pq}^\lambda \frac \partial {\partial \tilde f^b}
\bigr.
\nonumber\\
&&-\bigl.\Bigl[\frac 1{\sqrt{H}}\frac
\partial {\partial x^i}\left( \sqrt{H}h^{ni}\underset{\scriptscriptstyle{(\gamma)}}{{\mathscr A}^{\beta}_n}\right)  
+(G^{EC}\Lambda^{\beta}_E\Lambda^{\mu}_C)\frac 1{\sqrt{H}}
\frac {\partial (\sqrt{H}K^b_{\mu}\,)}{\partial  \tilde f^b}\Bigr](J_\beta )_{pq}^\lambda 
\nonumber\\
&&+({\gamma }^{\alpha \beta }+h^{ij}\underset{\scriptscriptstyle{(\gamma)}}{{\mathscr A}^{\alpha}_i}\underset{\scriptscriptstyle{(\gamma)}}{{\mathscr A}^{\beta}_j})(J_\alpha
)_{pq^{\prime }}^\lambda (J_\beta )_{q^{\prime }q}^\lambda \Bigr\}.
\label{dif_gen_ksi_res}
\end{eqnarray} 

This operator  acts in the space of the sections 
$\Gamma ({\tilde{\cal M}},V^{*})$ of the associated covector 
bundle 
${\cal E}^{*}={\tilde{\cal P}}\times _{\cal G}V^{\ast}_\lambda $ ($ {\tilde{\cal P}}=\cal P\times\cal \mathcal V$) with 
the following scalar product  
\begin{equation*}
(\varphi _p,\varphi _q)=\int_{\tilde{\cal M}}\langle \varphi _p,\varphi _q{\rangle}_
{V^{\ast}_{\lambda}}
 dv_{\tilde{\cal M}}(x,\tilde f),
\end{equation*}
 where $dv_{\tilde{\cal M}}(x,\tilde f)=\sqrt{H(x,\tilde f)}dx^1...dx^{n_{\cal M}}d\tilde f^1...\tilde f^{n_{\cal \mathcal V}}$. 

It is not difficult to show that the differential operator (\ref{dif_gen_ksi_res})  can also be expressed in terms of the horizontal Laplacian. Usually, for a  covector bundle, this operator is defined as follows:
 \begin{eqnarray*}
&&\!\!\!\!\!\left({\triangle}^{{\mathcal E^*}}\right)^{\lambda}_{pq}
={\sum}_{\bar{\scriptscriptstyle M}=1}^{n_{\cal P}}
\left({\nabla}^{\mathcal E^*}_{X^A_{\bar M}{\rm e_A}}
{\nabla}^{\mathcal E^*}_{X^B_{\bar M}{\rm e_B}}-
{\nabla}^{\mathcal E^*}_{{\nabla}^{\tilde{\mathcal M}}_{X^A_{\bar M}{\rm e_A}} {X^B_{\bar M}{\rm e_B}}}\right)^{\!\!\lambda}_{\!\!pq},\nonumber\\
&&\!\!\!\!\!\!
\end{eqnarray*}
in which ${X}^A_{\bar M}$  is a ``square root'' of the metric:
$\sum_{\bar{
{\scriptscriptstyle M}}
\scriptscriptstyle }
X_{\bar{M}}^AX_
{\bar{M}}^B=G^{AB}$ and the connection is  
$({\rm{\Gamma}}^{\mathcal E})^{\lambda}_{Bpq}={\mathscr A}^{\alpha}_B\,(J_{\alpha})^{\lambda}_{pq}$.

The covariant derivative  ${\nabla}^{{\pi}^*}$ is defined as 
\[
 {\nabla}^{{\pi}^*}_{{\rm e}_B}\varphi_p=\left({\rm I}^{\lambda}_{pq}\frac{\partial }{\partial Q^{*D}}-{\mathscr A}^{\alpha}_D(J_{\alpha})^{\lambda}_{pq}\right)\varphi_q,
\]
and 
\[
{\nabla}^{\scriptscriptstyle{\tilde{\mathcal M}}}_{\rm e_A}\, {\rm e_B}={}^{\tilde {\mathcal M}}{\Gamma}^C_{AB}\,{\rm e_C}.
\]
It should be noted that in our case we are dealing with  the product manifold, therefore, each capital index in the previous formulas means an abbreviation for two small indices.

Taking this into  account, one can find that the differential operator (\ref{dif_gen_ksi_res}) has the following representation:
\[
\frac12{\mu}^2{\kappa}\left[\bigl({\triangle}^{{\mathcal E^*}}\bigr)_{pq}^{\lambda}+{d}^{\mu \nu}
(J_\mu)_{pq^{\prime }}^\lambda 
 (J_\nu)_{q^{\prime }q}^\lambda \right]+\left(
\frac{1}{\mu ^2\kappa  m} {\tilde V}-\frac18\mu ^2\kappa  {\tilde J} \right)(I^\lambda )_{pq}\,.
\]

Also note, that the Green function ${\tilde G}^{\lambda}_{mn}$ satisfies the forward Kolmogorov equation with the operator 
\[
{\hat H}_{\kappa}=\frac{\hbar\kappa}{2m}\left[({\triangle}^{\mathcal E})^{\lambda}_{pq}+ {d}^{\mu \nu}
(J_\mu)_{pq^{\prime }}^\lambda 
 (J_\nu)_{q^{\prime }q}^\lambda \right]-\frac{\hbar \kappa}{8m}[\tilde J\,]I^\lambda _{pq}+
 \frac{\tilde V}{\hbar \kappa}I^\lambda _{pq},
\]
in which  the horizontal Laplacian $({\triangle}^{{\mathcal E}})^{\lambda}_{pq}$ is
\begin{eqnarray*}
&&\!\!\!\!\!\left({\triangle}^{{\mathcal E}}\right)^{\lambda}_{pq}
={\sum}_{\bar{\scriptscriptstyle M}}
\left({\nabla}^{\mathcal E}_{X^A_{\bar M}{\rm e_A}}
{\nabla}^{\mathcal E}_{X^B_{\bar M}{\rm e_B}}-
{\nabla}^{\mathcal E}_{{\nabla}^{\tilde{\mathcal M}}_{X^A_{\bar M}{\rm e_A}} {X^B_{\bar M}{\rm e_B}}}\right)^{\!\!\lambda}_{\!\!pq}\nonumber\\
\end{eqnarray*}

At $\kappa=i$ the forward Kolmogorov equation becomes the Schr\"odinger equation with the Hamilton operator $\hat H_{\mathcal E}=-\frac{\hbar}{\kappa}{\hat H}_{\kappa}|_{\kappa=i}$. The operator $\hat H_{\mathcal E}$ acts in the Hilbert space of the sections of the associated vector bundle ${\mathcal E}=\tilde{\mathcal P}\times_{\mathcal G}V_{\lambda}$.

\end{document}